\pdfoutput=1
\documentclass[conference]{IEEEtran}
\IEEEoverridecommandlockouts
\usepackage{cite}
\usepackage{amsmath,amssymb}
\usepackage{graphicx}
\usepackage{booktabs}
\usepackage{xcolor}
\usepackage{tikz}
\usetikzlibrary{positioning,arrows.meta,fit,calc,shapes.geometric}
\usepackage[hidelinks]{hyperref}
\hypersetup{pdftitle={WorldPager: Zero-Stall Tiered Memory for Long-Horizon World Models on a Single Consumer GPU},pdfauthor={Bingnan Liu, Ruini Xue}}
\graphicspath{{gen/}}

\newcommand{\sys}{WorldPager}
\newcommand{\Bc}{B_c}
\newcommand{\Tc}{T_c}

\begin{document}

\title{WorldPager: Zero-Stall Tiered Memory for Long-Horizon World Models on a Single Consumer GPU}

\author{\IEEEauthorblockN{Bingnan Liu}
\IEEEauthorblockA{School of Computer Science and Engineering\\
University of Electronic Science and Technology of China\\
Chengdu, China\\
2024080401011@std.uestc.edu.cn}
\and
\IEEEauthorblockN{Ruini Xue\thanks{Corresponding author.}}
\IEEEauthorblockA{School of Computer Science and Engineering\\
University of Electronic Science and Technology of China\\
Chengdu, China\\
xueruini@uestc.edu.cn}}

\maketitle

\begin{abstract}
Interactive world models keep the KV of past chunks as retrieval memory so that revisited places stay consistent. On a single consumer machine (one RTX~4090, 32\,GiB of host memory, one NVMe SSD), long sessions with such memory hit two walls. The first is capacity: the KV of Matrix-Game~2.0 exhausts host memory after 45 seconds of video in bf16 and after 2.8 minutes in int4. The second is the model's positional horizon of 4.3 minutes. We show that this memory is capacity-bound rather than bandwidth-bound and that its access set is known before compute, because the camera pose is an input. An analytical model built on these observations predicts spill points to within one chunk and classifies zero versus non-zero stalls correctly on held-out configurations. Guided by it, WorldPager stores int4 KV in a GPU/DRAM/NVMe hierarchy instead of recomputing it, prefetches each chunk's memory from the extrapolated pose, never waits on a miss, and writes back in the background; to our knowledge, it is the first to place the KV memory of an interactive world model on an NVMe tier with zero-stall prefetching. With training-free position handling, it generates 37.5 minutes of video continuously with zero stalls and constant host memory, at about 9.5 frames per second. On a sealed test set, int4 storage shows no measurable quality loss; on long-gap revisits it lowers image quality by about one MUSIQ point.
\end{abstract}

\section{Introduction}\label{sec:intro}

Interactive world models generate video chunk by chunk in response to a user's keyboard and mouse actions~\cite{matrixgame2,lingbot2,worldplay}.
For such a world to stay consistent, a place the user returns to must look as it did before, and recent systems achieve this with retrieval memory: they keep the keys and values (KV) of past chunks and let a new chunk attend to the ones whose camera pose matches~\cite{worldkv,reworld}.
These systems are evaluated on data-center GPUs.
We study the same workload under tight resources: one consumer GPU (an RTX~4090 with 24\,GB), a 32\,GiB host-memory budget, and one NVMe SSD.
The question is how long such a machine can keep generating a consistent world.

\begin{figure}[t]
\centering
\includegraphics[width=\columnwidth]{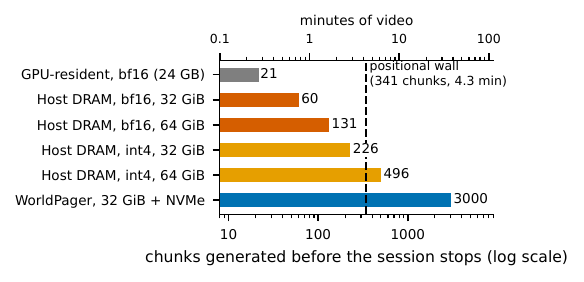}
\caption{Two walls for MG2 on one RTX 4090. Capacity stops every design without an NVMe tier early (chunk 21, 60, or 226 at 32\,GiB), and without position handling every design stops by the positional wall at chunk 341; \sys{} reaches 3000 chunks. All bars are measured; the 64\,GiB int4 run uses an extended position table so that it isolates the capacity wall, and it stops at chunk 496 against a registered model prediction of 492. \sys{} runs under 32\,GiB with an NVMe tier (37.5\,min of video).}
\label{fig:walls}
\end{figure}

Two walls stop it, and Fig.~\ref{fig:walls} shows both for Matrix-Game~2.0 (MG2)~\cite{matrixgame2}.
The first wall is \emph{capacity}.
The KV of one chunk, 0.75\,s of video, occupies 486.6\,MB in bf16, and retrieval memory keeps all of it.
Kept on the GPU, memory runs out at chunk 21 (16\,s of video); kept in host memory, at chunk 60 (45\,s); quantized to int4 and kept in host memory, at chunk 226 (2.8 minutes).
The second wall is the \emph{positional horizon}: the model's rotary position table covers 1024 latent frames, and generation fails at chunk 341 (4.3 minutes) whatever the memory system.

Existing approaches do not remove these walls on a single consumer machine.
LLM offloading systems move KV between GPU, host memory, and disk~\cite{flexgen,infinigen,cachedattention}, but they are built on the premise that decoding is limited by the bandwidth of that movement, and those that load only a subset of entries must compute or speculate the access set at run time~\cite{infinigen}.
World-model memory systems keep KV on the GPU~\cite{worldkv} or in host memory~\cite{reworld,omnimem,inferix}, which only moves the first wall from chunk 21 to chunk 60; WorldKV notes that transfer latency limits its CPU-offloaded variant~\cite{worldkv}.
WorldPlay~\cite{worldplay} keeps frames and recomputes their KV for every chunk, which costs 3.07$\times$ a denoising step per retrieved chunk on our GPU.
KV quantization~\cite{kivi,quantwm,qvg} reduces bytes per chunk by a constant factor, which delays the first wall but does not remove it.

Our starting point is a characterization of this workload.
World-model memory is \emph{capacity-bound, not bandwidth-bound}: a chunk takes about 1.2\,s to compute, while the KV it must read fits in that window many times over, even from NVMe.
Its access set is also \emph{known before compute}: the retrieval key is the camera pose, which follows from the control input.
An operating system cannot prefetch reliably because the future page-reference string is unknown; here the control input makes that string known one chunk ahead.
From these two properties we derive a first-order analytical model that predicts where each budget spills, how much bandwidth prefetching needs, and how long an on-demand design stalls, and we test it on configurations held out from its construction.

Guided by the model, we build \sys{}, a memory runtime that sits below the retrieval policy.
It stores int4 KV in a GPU / pinned-DRAM / NVMe hierarchy instead of recomputing it, prefetches each chunk's retrieval set from the extrapolated pose while the previous chunk is denoised, falls back to the previous set on a miss instead of waiting, and writes spilled blocks back in the background.
To cross the second wall it handles positions at inference time, without training: the position table is extended, which leaves the native horizon bit-identical, and retrieved memory is re-placed near the current frame beyond it, following established practice~\cite{streamingllm,rollingforcing,deepforcing,worldplay}.

We make three contributions.
\begin{itemize}
\item \textbf{Two walls, characterized and modeled.} We show that memory-augmented world models on a consumer machine hit a capacity wall long before a bandwidth limit, and a positional wall at the model's native horizon. Our analytical model predicts the spill point, the bandwidth at which prefetching fails, and the stall of on-demand swap-in. On held-out configurations its eight spill-point and first-read predictions fall within one chunk, all 15 zero/non-zero stall predictions are correct, and 22 of 24 registered criteria on MG2 and int8 are met (Sec.~\ref{sec:char}).
\item \textbf{Crossing the first wall.} We store int4 KV instead of recomputing it and, to our knowledge, \sys{} is the first system to place the KV memory of an interactive world model on an NVMe tier with zero-stall prefetching. With all retrieved KV read from NVMe, prefetching removes the 60\,ms per-chunk stall of on-demand swap-in and keeps the median stall at zero down to 0.25\,GB/s (Sec.~\ref{sec:design}, Sec.~\ref{sec:eval-nvme}).
\item \textbf{Crossing the second wall, end to end.} With training-free position handling, one consumer machine generates 37.5 minutes of video continuously, instead of 45 seconds before host memory runs out (2.8 minutes with int4 alone), with zero stalls and constant host memory. Long-gap memory is read back from NVMe without stalls; int4 storage leaves revisit consistency unharmed and lowers long-gap image quality by about one MUSIQ point; on a sealed test set, int4 shows no measurable quality loss (Sec.~\ref{sec:eval-long}, Sec.~\ref{sec:eval-quality}).
\end{itemize}
\sys{} does not make generation real-time: the 37.5 minutes of video take 63 minutes to generate (about 9.5\,fps).
Sec.~\ref{sec:analysis} reports what did not work and why, and Sec.~\ref{sec:concl} lists the limits of our evidence.

\section{Two Walls: Characterization and an Analytical Model}\label{sec:char}

This section characterizes memory-augmented generation on our platform, identifies the two walls that bound a session, and turns the characterization into an analytical model whose predictions we test on configurations that were not used to build it.

\subsection{Setting}\label{sec:char-setting}

A chunk-wise interactive world model generates video one chunk at a time, conditioned on the user's keyboard and mouse actions.
MG2~\cite{matrixgame2} has 1.8B parameters including its action modules; a chunk is three latent frames, or 12 video frames (0.75\,s at 16\,fps), and is denoised in four steps.
Attention covers an attention sink, a short local window, and, with retrieval memory, $k{=}3$ earlier chunks selected by camera pose~\cite{worldkv,reworld}.
Retrieval memory therefore keeps the KV of every generated chunk, and that KV grows linearly with session length: one chunk occupies $B_c = 486.6$\,MB in bf16 across the 30 layers.
Our platform is one RTX 4090 (24\,GB), a 32\,GiB host-memory budget, and one NVMe SSD (5.82\,GB/s sequential read with \texttt{O\_DIRECT}; 24--25\,GB/s host-to-device copies from pinned memory).

\subsection{The first wall: capacity, not bandwidth}\label{sec:char-capacity}

Memory runs out long before moving it becomes a problem.
Keeping retrieved KV on the GPU exhausts the 24\,GB at chunk 21 (16\,s of video). Counting allocated bytes alone gives an upper bound of 28 chunks; adding the allocator slack and non-PyTorch memory measured in the same run gives 21, which is a calibration rather than an independent prediction.
Moving it to host memory in bf16 lasts until chunk 60 (45\,s), and int4 storage until chunk 226 (2.8\,min); all three points come before the model's native limit of 341 chunks.

Bandwidth, by contrast, is not the constraint.
A chunk takes about 1.2\,s to compute, and reading 1.46\,GB (the KV of three bf16 chunks) from NVMe into pinned memory and on to the GPU takes 276\,ms when the two stages are pipelined.
With one chunk of double buffering, the read is hidden completely: in a microbenchmark with the real MG2 computation and synthetic transfer payloads, the stall is 0.0\,ms for 0.37\,GB and 1.46\,GB payloads from DRAM and from NVMe, against 15.8 to 289.1\,ms when the same payloads are copied synchronously.
The working set per chunk is small compared with the compute window, while the total state grows without bound.

This is the opposite of the premise of LLM offloading.
Systems such as FlexGen~\cite{flexgen} and InfiniGen~\cite{infinigen} offload KV because decoding is limited by the bandwidth of moving it: a decoding step is short, so a transfer cannot hide behind it.
Table~\ref{tab:premise} summarizes the contrast.
We use this observation as the basis of the design; we do not claim the general observation that video diffusion is compute-heavy, which is well known.

\begin{table}[t]
\centering
\caption{Premises of KV offloading in LLM decoding versus world-model memory (qualitative; our measurements for the right column).}
\label{tab:premise}
\footnotesize
\resizebox{\columnwidth}{!}{%
\begin{tabular}{@{}lll@{}}
\toprule
 & LLM decoding (selective KV) & World model with retrieval memory \\
\midrule
Compute per step & short (one token) & long ($\approx$1.2\,s per chunk) \\
Binding limit & transfer bandwidth & capacity (chunk 21 / 60 / 226) \\
Access set known & after the query is computed & before compute (pose is an input) \\
Prefetch & speculative~\cite{infinigen} & from the control input \\
\bottomrule
\end{tabular}}
\end{table}

\subsection{The access set is known before compute}\label{sec:char-access}

In LLM decoding with selective KV retrieval, the entries a step needs depend on its query, which exists only after the preceding computation; systems either select them on the critical path or speculate, as InfiniGen does by rehearsing the next layer's attention~\cite{infinigen}.
(Coarser look-ahead exists, for example a fixed layer order~\cite{flexgen} or a queue of waiting requests~\cite{cachedattention}.)
In an interactive world model, the retrieval key is the camera pose, which follows from the user's actions.
An operating system cannot prefetch reliably because the future page-reference string is unknown; here the control input makes the reference string known one chunk ahead.
In practice the next action may not be known yet, so \sys{} extrapolates the pose, and the prediction can miss.
On synthetic trajectories of four motion classes, constant-velocity extrapolation finds the retrieval set of the next chunk with a hit rate of 0.66--0.94, against 0.03--0.52 when the current pose is reused (CPU simulation, 50 trajectories of 800 chunks per class, assuming a 90\textdegree{} field of view).
In the end-to-end runs, the fraction of retrieved blocks that the prefetch missed is 0.010 on a\_yaw and 0.020 on b1\_ws, and the stale-set fallback covers these cases.

\subsection{The second wall: the positional horizon}\label{sec:char-rope}

MG2 encodes time with a rotary position table of 1024 latent frames, so with the original table generation fails at chunk 341 (about 4.3 minutes of video).
This wall is independent of memory.
Removing it by recomputing memory at new positions, as WorldPlay does~\cite{worldplay}, costs 3.07$\times$ a denoising step per retrieved chunk on our GPU (Sec.~\ref{sec:design-store}); Sec.~\ref{sec:design-rope} handles it at inference time instead.

\subsection{An analytical model}\label{sec:char-model}

We turn the characterization into five first-order relations.
Let $B_c$ be the bytes stored per chunk, $T_c$ the compute time of a chunk (the window available to hide a transfer), $P$ the host-memory pool (the budget minus the process baseline and a 2\,GiB margin), $r_{\mathrm{new}}$ the bytes a chunk must read from NVMe because they are not yet staged, and $\mathit{BW}_R$, $\mathit{BW}_W$ the NVMe read and write bandwidths.
\begin{align}
N_{\mathrm{fit}} &= 1 + \lfloor P / B_c \rfloor \label{eq:spill}\\
\text{stall}_{\mathrm{prefetch}} &\approx \max\!\left(0,\; r_{\mathrm{new}}/\mathit{BW}_R - T_c\right) \label{eq:prefetch}\\
\text{stall}_{\mathrm{on\text{-}demand}} &\approx r_{\mathrm{new}}/\mathit{BW}_R + o \label{eq:sync}\\
\mathit{BW}_W &\ge B_c / T_c \label{eq:write}\\
\text{NVMe use} &\ge (N - N_{\mathrm{fit}}) \cdot B_c \label{eq:usage}
\end{align}
Eq.~\eqref{eq:spill} gives the spill point, the chunk at which a DRAM-only design runs out of host memory and a tiered design starts writing to NVMe (the 1 is the attention sink).
Eq.~\eqref{eq:prefetch} gives the stall of prefetching and, by setting it to zero, the bandwidth below which prefetching fails, $r_{\mathrm{new}}/T_c$.
Eq.~\eqref{eq:sync} gives the stall of on-demand swap-in, which cannot overlap with compute; $o$ is a fixed per-chunk overhead of submitting 30 per-layer reads and copies, which we fit (about 34\,ms) rather than measure.
Eq.~\eqref{eq:write} gives the write bandwidth that background write-back needs, and Eq.~\eqref{eq:usage} a lower bound on NVMe usage for a session of $N$ chunks.
For MG2 with int4 storage, $B_c = 129.4$\,MB, $T_c \approx 1.25$\,s, and in steady state a chunk reads about one new chunk ($r_{\mathrm{new}} \approx 121.5$\,MB on a\_yaw); the model therefore predicts that prefetching fails near 0.1\,GB/s and that write-back needs about 0.11\,GB/s, two orders of magnitude below what the drive provides.
The model is first order: it ignores the variance of $r_{\mathrm{new}}$ across chunks and assumes that the drive delivers its nominal bandwidth.
It explains the design choices of Sec.~\ref{sec:design} (Table~\ref{tab:design-map}); it does not derive them.

\textbf{Consistency on the data used to build it.}
With the pool size of each run, Eq.~\eqref{eq:spill} reproduces all nine measured capacity boundaries exactly, including the walls at chunks 60 and 226 under 32\,GiB and 131 under 64\,GiB.
We report this only as consistency, because these runs informed the model.

\textbf{Held-out validation.}
We then registered predictions for configurations that were not used to build the model and ran them without adjusting the predictions (Table~\ref{tab:validation}).
All eight spill-point and first-NVMe-read predictions, from four held-out configurations, fall within one chunk of the measurement: int8 storage (emulated by padding int4 content to the int8 block size, so it tests byte accounting only, not int8 quality or dequantization), a 64\,GiB round trip on MG2, and LingBot-World~\cite{lingbot2} round trips of 200 and 240 chunks.
All 15 zero/non-zero stall predictions are correct, and 7 of the 9 non-zero stall magnitudes are within 30\%; overall, 22 of the 24 registered criteria on MG2 and int8 are met.
On LingBot-World, 45 of 48 read-path criteria are met, and the SATA-class emulation of Sec.~\ref{sec:eval-nvme} meets all 10.

We describe three misses.
(i) On b1\_ws at 0.1\,GB/s, just below the predicted failure bandwidth of 0.103\,GB/s, prefetching stalls 75.7\,ms against a predicted 38.1\,ms; near the transition the two terms of Eq.~\eqref{eq:prefetch} nearly cancel, so small errors are amplified: a limiter that delivers about 3.7\% less than its nominal rate adds about 50\,ms, and a compute window about 11\,ms longer than assumed removes part of it.
(ii) On LingBot-World at 0.075\,GB/s the model overestimates the stall by 71\% (448.5 against 262.6\,ms, three repetitions), a systematic error that we attribute to the choice of $T_c$ and to the limiter's accounting.
(iii) On the 240-chunk LingBot-World round trip, the measured volume read back from NVMe is 26.6\% below the prediction (2.20 against 3.0\,GiB), because the model overestimates how many blocks are read back.
We later derived two corrections (a read-back term in the write volume and a per-chunk rather than mean read size near the failure bandwidth); they were fitted after the fact and are not validated.

\begin{table}[t]
\centering
\caption{Held-out validation of the analytical model: all spill-point and zero/non-zero stall predictions hold, and 22 of 24 MG2 criteria are met. Predictions were registered before the runs and not adjusted.}
\label{tab:validation}
\footnotesize
\resizebox{\columnwidth}{!}{%
\begin{tabular}{@{}llc@{}}
\toprule
Prediction & Held-out configurations & Met \\
\midrule
Spill point and first NVMe read ($\pm$1 chunk) & int8 (emulated), MG2 64\,GiB, LingBot 200 / 240 & 8 / 8 \\
Zero vs.\ non-zero stall & b1\_ws, int8, LingBot (rate-limited) & 15 / 15 \\
Non-zero stall within 30\% & same & 7 / 9 \\
All registered criteria (MG2, int8) & & 22 / 24 \\
LingBot read path & 200 / 240 chunks, 3 repetitions & 45 / 48 \\
SATA-class emulation (0.5\,GB/s) & MG2, 2000 chunks, 2 repetitions & 10 / 10 \\
\bottomrule
\end{tabular}}
\end{table}

\textbf{Projections.}
Used as a design guide, the model indicates what other budgets need (values are projections unless noted).
With 64\,GiB of host memory, MG2 with int4 fills the pool at chunk 492, about 10 minutes of wall-clock time, so a one-hour session still needs a storage tier of roughly 300--600\,GiB. We later checked this point directly: a DRAM-only int4 run with 64\,GiB, registered with a prediction of 492 (window 477--507) before it ran, stopped at chunk 496 (+0.8\%). Its registered tolerance was $\pm$3\%, wider than the one-chunk criterion of the spill predictions, so it is not counted among the eight in Table~\ref{tab:validation}.
Prefetching needs about 0.10\,GB/s of read bandwidth and write-back about 0.11\,GB/s (up to 0.21\,GB/s with read-back), so even a SATA SSD, whose vendor specification is about 0.5\,GB/s, would suffice by bandwidth.
For LingBot-World the corresponding figures are 219 chunks (about 12.7 minutes), 0.085\,GB/s, and 0.085\,GB/s.
For a storage tier, the requirement is capacity, not bandwidth.

\section{WorldPager Design}\label{sec:design}

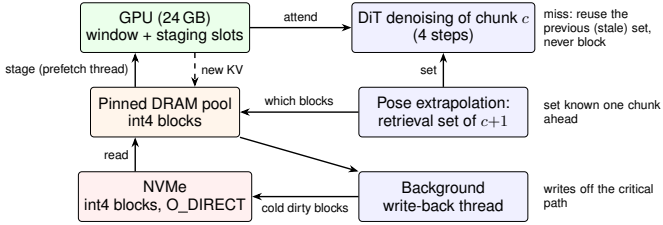
\begin{figure}[t]
\centering
\resizebox{\columnwidth}{!}{%
\begin{tikzpicture}[
  font=\sffamily\small,
  tier/.style={draw, rounded corners=2pt, minimum width=3.0cm, minimum height=1.0cm, align=center},
  proc/.style={draw, rounded corners=2pt, fill=blue!6, align=center, minimum width=3.4cm, minimum height=1.0cm},
  note/.style={font=\sffamily\scriptsize, align=left, anchor=west, text width=2.4cm},
  arr/.style={-{Stealth[length=2mm]}, thick},
  darr/.style={-{Stealth[length=2mm]}, thick, dashed}
]
\node[tier, fill=green!8]  (gpu)  at (0,3.4) {GPU (24\,GB)\\window + staging slots};
\node[tier, fill=orange!8] (dram) at (0,1.7) {Pinned DRAM pool\\int4 blocks};
\node[tier, fill=red!6]    (nvme) at (0,0.0) {NVMe\\int4 blocks, O\_DIRECT};
\node[proc] (dit)  at (5.6,3.4) {DiT denoising of chunk $c$\\(4 steps)};
\node[proc] (pred) at (5.6,1.7) {Pose extrapolation:\\retrieval set of $c{+}1$};
\node[proc] (bgw)  at (5.6,0.0) {Background\\write-back thread};
\draw[arr] (gpu) -- node[above, font=\sffamily\scriptsize]{attend} (dit);
\draw[arr] (pred) -- node[left, font=\sffamily\scriptsize]{set} (dit);
\draw[arr] (pred) -- node[above, font=\sffamily\scriptsize]{which blocks} (dram);
\draw[arr] ([xshift=-6mm]nvme.north) -- node[left, font=\sffamily\scriptsize]{read} ([xshift=-6mm]dram.south);
\draw[arr] ([xshift=-6mm]dram.north) -- node[left, font=\sffamily\scriptsize]{stage (prefetch thread)} ([xshift=-6mm]gpu.south);
\draw[darr] ([xshift=6mm]gpu.south) -- node[right, font=\sffamily\scriptsize]{new KV} ([xshift=6mm]dram.north);
\draw[arr] (dram.south east) -- (bgw.north west);
\draw[arr] (bgw) -- node[below, font=\sffamily\scriptsize]{cold dirty blocks} (nvme);
\node[note] at (7.5,3.4) {miss: reuse the previous (stale) set, never block};
\node[note] at (7.5,1.7) {set known one chunk ahead};
\node[note] at (7.5,0.0) {writes off the critical path};
\end{tikzpicture}}
\caption{\sys{} overview. KV of past chunks is stored as int4 blocks (one block per layer and chunk) in a pinned DRAM pool and spills to NVMe. While chunk $c$ is denoised, the retrieval set of chunk $c{+}1$ is computed from the extrapolated camera pose and staged to the GPU by a prefetch thread. A miss reuses the previous retrieval set instead of waiting. Cold dirty blocks are written to NVMe by a background thread.}
\label{fig:overview}
\end{figure}

\sys{} is a memory runtime that sits below the retrieval policy of a chunk-wise world model.
The policy decides \emph{which} past chunks a new chunk attends to; \sys{} decides \emph{where} their keys and values live and \emph{when} they move.
Each design choice follows from one term of the analytical model in Sec.~\ref{sec:char}.
Fig.~\ref{fig:overview} gives the overview, and Table~\ref{tab:design-map} maps the components to the model terms they answer.

\begin{table}[t]
\centering
\caption{Design choices and the model terms they answer (Sec.~\ref{sec:char}).}
\label{tab:design-map}
\footnotesize
\begin{tabular}{@{}p{0.43\columnwidth}p{0.49\columnwidth}@{}}
\toprule
Component & Model term it answers \\
\midrule
Store int4 KV, not latents (\ref{sec:design-store}) & per-chunk bytes $\Bc$; recompute cost vs.\ $\Tc$ \\
Three-tier store (\ref{sec:design-tiers}) & spill point $1+\lfloor P/\Bc\rfloor$ \\
Pose-driven prefetch, stale fallback (\ref{sec:design-prefetch}) & zero-stall condition $r_{\mathrm{new}}/\mathit{BW}_R \le \Tc$ \\
Background write-back (\ref{sec:design-write}) & write bandwidth $\ge \Bc/\Tc$ \\
Position handling (\ref{sec:design-rope}) & native horizon of 341 chunks \\
\bottomrule
\end{tabular}
\end{table}

\subsection{Store quantized KV instead of recomputing it}\label{sec:design-store}

A memory system for a world model can keep either the generated latent frames or the keys and values computed from them.
WorldPlay~\cite{worldplay} keeps frames: for every chunk it rebuilds a memory context and recomputes the memory KV from it (its Algorithm~2 resets the cache from the reconstituted context), and it reports real-time inference on eight H800 GPUs.
On our platform this choice does not fit the time budget.
Recomputing the KV of one retrieved chunk of Matrix-Game~2.0 (MG2)~\cite{matrixgame2} from its clean latents, which replays the chunk and its two predecessors through all 30 layers, takes 585.5\,ms on an RTX~4090, or 3.07$\times$ one denoising step (190.4\,ms).
A chunk retrieves three such chunks and is denoised in four steps, so recomputation would more than double the chunk time.

\sys{} therefore stores the KV.
The cost of this choice is capacity: one chunk of MG2 KV occupies 486.6\,MB in bf16, which is the quantity that exhausts host memory in Sec.~\ref{sec:char}.
We reduce it with standard KV quantization.
Keys are quantized per channel and values per token, following KIVI~\cite{kivi}; we use symmetric int4 with groups of 64 and bf16 scales, which brings a chunk to 129.4\,MB (26.6\% of bf16).
Only the remote chunks that the retrieval policy may read are quantized; the attention sink, the local window, and the chunk being generated stay in bf16.
The bit width was fixed by a rule registered before the bit-width sweep: the lowest width whose revisit-retention confidence bound stays above 95\% and whose image quality does not drop measurably, which selected int4 over int3 and int2 (Sec.~\ref{sec:analysis}).
Reading a block back costs only dequantization, 10.9\,ms per chunk in the median, instead of a 585.5\,ms recomputation per retrieved chunk.
Under the same 32\,GB host budget, the pool holds 226 int4 chunks against 60 bf16 chunks (3.8$\times$).
We do not claim novelty for the quantizer.
QuantWM~\cite{quantwm} and Quant VideoGen~\cite{qvg} report 2-bit KV quantization for world and video models, and \sys{} can use such a quantizer unchanged; Sec.~\ref{sec:analysis} reports what 2 bits cost on our setting.

\subsection{A three-tier KV store}\label{sec:design-tiers}

The store has three tiers (Fig.~\ref{fig:overview}).
The GPU holds the attention window and a set of staging slots for the retrieved chunks.
A pinned host-memory pool holds int4 blocks; the pool is sized to the host budget minus the process baseline and a 2\,GiB margin (27.2\,GiB under a 32\,GB cgroup limit).
An NVMe file holds the blocks that do not fit, accessed with \texttt{O\_DIRECT} so that NVMe traffic does not occupy the page cache, which would count against the same host budget.
The unit of storage and transfer is one block per layer and chunk (about 4.3\,MB in int4), aligned to 4\,KiB.
A block is written once to NVMe; a block read back from NVMe stays clean, so evicting it later costs no write.

\subsection{Pose-driven prefetch with a stale-set fallback}\label{sec:design-prefetch}

In an interactive world model, the retrieval key of the next chunk is its camera pose, which the controller provides before the chunk is generated.
\sys{} uses the access property of Sec.~\ref{sec:char-access} as follows.
While chunk $c$ is denoised, a prefetch thread extrapolates the camera pose of chunk $c{+}1$ at constant velocity, evaluates the retrieval policy on the extrapolated pose, and stages the selected blocks from DRAM or NVMe into the GPU staging slots.
Blocks already staged for chunk $c$ are reused, so in steady state a chunk usually moves only the blocks that are new to its retrieval set.
When chunk $c{+}1$ starts, it attends to the blocks that the policy selects at its actual pose if they are staged.
If some are not, the chunk does not wait: it reuses the retrieval set staged for chunk $c$, which we call the \emph{stale-set fallback}.
For example, if a revisit begins at chunk 240 and the extrapolated pose of chunk 240 misses it, chunk 240 attends to the blocks staged for chunk 239 instead of waiting for the revisited blocks.
The fallback trades a possible one-chunk delay of memory for a guarantee that generation never waits on storage; Sec.~\ref{sec:char} reports the measured miss ratio.
Dequantization runs on the compute stream after staging and is not overlapped with the previous chunk.

Prefetching world-model memory from host memory while denoising is not new: ReWorld~\cite{reworld} keeps a bounded landmark bank in pinned host memory and prefetches the landmarks nearest to the \emph{current} pose on a separate stream.
\sys{} differs in three respects: it selects blocks for the \emph{next} chunk from an extrapolated pose, it never blocks on a miss (OmniMem~\cite{omnimem} waits until a selected block is resident, and HERALD~\cite{herald} reports a stall at block boundaries when retrieval runs late), and it extends the store to an NVMe tier with an unbounded number of chunks.

\subsection{Write path and bookkeeping}\label{sec:design-write}

Once the pool is full, every new chunk must evict an old one.
Writing the victim synchronously puts an NVMe write on the critical path of every chunk (Sec.~\ref{sec:eval-write}).
\sys{} instead keeps new blocks in a dirty queue in arrival order, which approximates coldest first.
When the reclaimable space in the pool (free space, clean blocks, and blocks already queued) falls below three chunks, a background thread writes the coldest dirty blocks to NVMe with double buffering, in the same way that CachedAttention saves KV asynchronously~\cite{cachedattention}.
A block that is being written stays in the pool and is served from memory if it is read.
Allocation then evicts the least recently used clean block.
Two implementation details keep the per-chunk bookkeeping independent of session length.
Retrieval-set selection is vectorized; at chunk 2900 it takes 2.12\,ms instead of 38.4\,ms, with identical selections over 3000 chunks.
Victim selection uses a lazily updated heap instead of a scan over all blocks; on the return leg of a 2000-chunk round trip this reduces the median write-back time per chunk from 105.4\,ms to 27.6\,ms, with a bit-identical video.

\subsection{Beyond the native horizon: inference-time position handling}\label{sec:design-rope}

With the original position table, MG2 fails at chunk 341 (the second wall, Sec.~\ref{sec:char-rope}).
\sys{} handles it at inference time and without training, in two parts.
First, it extends the position table (to 8192 rows, or 12288 for the 3000-chunk session).
Within the native horizon this changes nothing: for every chunk index below 341 the output is bit-identical to the original model.

Second, beyond the native horizon, retrieved memory would sit at relative distances never seen in training.
Because the cache holds keys after rotation, \sys{} re-rotates each retrieved key by the difference between its stored position and a target position just before the window: the most recent retrieved chunk is placed 30 frames before the current chunk, older retrieved chunks and the attention sink are placed contiguously before it, and only the 44 temporal channels of each head are rotated.
Values and the window are unchanged.
The rule is hybrid: it is disabled while the chunk index is below 341, so that the first 341 chunks remain bit-identical to the model with the extended table, and it is enabled from chunk 341 on.
We verified both properties on a 24-chunk development run and a 400-chunk round trip, where the first differing chunk is chunk 341.
The re-rotation adds about 27\,ms per chunk in paired round-trip runs; in the 3000-chunk session the chunk time steps up by 21.3\,ms at chunk 341.

Re-placing cached content at positions close to the current frame is established practice.
StreamingLLM assigns positions within the cache rather than in the original text~\cite{streamingllm}; Rolling Forcing~\cite{rollingforcing} and Deep Forcing~\cite{deepforcing} relocate the temporal positions of attention-sink frames for long video generation; and WorldPlay reassigns the positions of all retrieved context frames so that they stay close to the current frame~\cite{worldplay}.
We do not claim the idea.
We apply it, as Rolling Forcing does, by rotating cached keys; what is specific here is only the setting: the shifted keys are pose-retrieved long-term memory that may come back from NVMe, and the shift is enabled only beyond the native horizon so that the first 341 chunks stay bit-identical.
Its effect on quality depends on the scene (Sec.~\ref{sec:analysis}); we treat it as an optional component that is needed only beyond 341 chunks.

\section{Evaluation}\label{sec:eval}

We ask five questions.
(Q1) What does \sys{} cost per chunk while memory still fits in DRAM?
(Q2) Does prefetching keep the NVMe tier off the critical path, and how does it compare with the on-demand swap-in used by LLM serving systems?
(Q3) What does spilling to NVMe cost on the write path?
(Q4) Does a session run past both walls without stalls or memory growth, and is long-gap memory recalled from NVMe without stalls and without loss from int4 storage?
(Q5) Is quality preserved, and do the system results hold on a second model?

\subsection{Setup}\label{sec:eval-setup}

\textbf{Platform.}
One RTX 4090 (24\,GB), one NVMe SSD, and a host-memory budget of 32\,GiB (64\,GiB where noted), enforced with a cgroup limit without swap.
NVMe accesses use \texttt{O\_DIRECT}.
Bandwidth limits are imposed by a software token bucket on the NVMe path; they emulate slower devices and are not measurements of real ones.

\textbf{Model and workload.}
MG2~\cite{matrixgame2} (1.8B parameters including action modules) generates chunks of 12 frames (0.75\,s of video at 16\,fps) from keyboard and mouse actions.
The retrieval policy selects $k{=}3$ past chunks by camera pose, as in WorldKV~\cite{worldkv}.
Timing uses two synthetic trajectories with dense revisits, a yaw sweep (a\_yaw) and a walk-and-strafe path (b1\_ws), each 36 chunks long, plus longer round trips described where used.
Each timing configuration runs five rollouts after one discarded warm-up rollout, giving 165 steady-state chunks.
All timing comes from one starting image and one seed, and we do not report confidence intervals for it.
Host budgets are cgroup memory limits in GiB, and pinned memory counts against them.
A \emph{stall} is the time a chunk waits for its retrieved KV; it excludes dequantization and write-back. We count a chunk as stalled when it waits more than 0.5\,ms, and ``zero stalls'' means that no chunk does.

\textbf{Baselines} (all are our re-implementations on the same model and retrieval policy).
\emph{Sliding window}: no retrieval memory.
\emph{WorldKV-style}: retrieved KV kept on the GPU in bf16; the official MG2 code is not public, so we reuse our implementation of its retrieval rule.
\emph{ReWorld-style}: all chunks in bf16 in host DRAM, prefetched one chunk ahead from the current pose, as described for ReWorld~\cite{reworld}, to which we add \sys{}'s stale fallback; unlike ReWorld, it neither caps the bank nor trains the model.
\emph{Synchronous}: retrieved chunks are copied when the chunk starts.
\emph{On-demand swap-in}: the synchronous copy with an LRU staging buffer on the GPU, so that a block staged for the previous chunk is reused.
It corresponds to the swap-in path of LLM serving systems that move KV on demand~\cite{flexgen,cachedattention} and is the fair reference for prefetching.

\subsection{Per-chunk cost below the first wall (Q1)}\label{sec:eval-e2e}

Table~\ref{tab:e5} reports the 36-chunk runs under 32\,GiB, where every method still fits.
\sys{} takes 1210.5\,ms per chunk (9.91\,fps) with zero stalls, against 1022.9\,ms for the sliding window, which reads no memory.
At this scale the prefetching designs and synchronous int4 are within 2\% of each other (the synchronous int4 copy stalls 15.8\,ms per chunk); synchronous bf16 is 4.3\% slower than the ReWorld-style design because of its 58.2\,ms stall.
The differences that matter appear at the walls: the WorldKV-style design runs out of GPU memory at chunk 21, both bf16 host-memory designs exhaust host memory at chunk 60, and int4 without an NVMe tier at chunk 226.

\begin{table}[t]
\centering
\caption{Per-chunk cost at 32\,GiB (a\_yaw, 36 chunks; median / P99 over 165 chunks) and the chunk at which each design exhausts memory.}
\label{tab:e5}
\footnotesize
\setlength{\tabcolsep}{2.5pt}
\resizebox{\columnwidth}{!}{%
\begin{tabular}{@{}lcccc@{}}
\toprule
Design (storage) & Chunk (ms) & FPS & Stall (ms) & Exhausted \\
\midrule
Sliding window (none) & 1022.9 / 1029.3 & 11.73 & 0 & -- \\
WorldKV-style (GPU, bf16) & -- & -- & -- & 21 (GPU) \\
Synchronous (DRAM, bf16) & 1259.0 / 1275.2 & 9.53 & 58.2 & 60 (host) \\
ReWorld-style (DRAM, bf16) & 1207.6 / 1223.4 & 9.94 & 0 & 60 (host) \\
Synchronous (DRAM, int4) & 1230.8 / 1253.2 & 9.75 & 15.8 & 226 (host) \\
\textbf{\sys{}} (tiers, int4) & 1210.5 / 1226.2 & 9.91 & 0 & not reached \\
\bottomrule
\end{tabular}}
\end{table}

\subsection{The NVMe tier: prefetch versus on-demand swap-in (Q2)}\label{sec:eval-nvme}

To isolate the NVMe tier we force every retrieved block to come from NVMe.
Without a bandwidth limit, \sys{} takes 1255.8\,ms per chunk with zero stalls on a\_yaw, while on-demand swap-in stalls 60.2\,ms (P50) and takes 1309.5\,ms, 4.3\% longer.
A synchronous copy without staging reuse stalls 111.6\,ms; part of that gap comes from reuse rather than from prefetching, so we use on-demand swap-in as the reference.
b1\_ws gives the same picture (0 against 61.2\,ms).

\begin{figure}[t]
\centering
\includegraphics[width=\columnwidth]{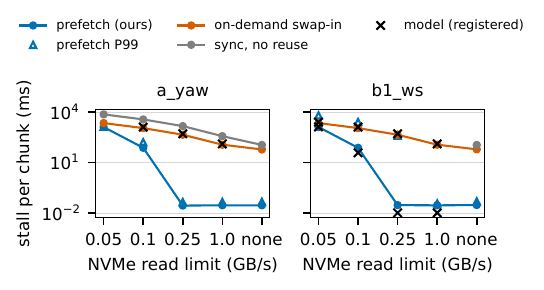}
\caption{Prefetching keeps the median stall at zero down to 0.25\,GB/s, whereas on-demand swap-in stalls at every bandwidth. Stall per chunk (P50; triangles: P99 of prefetching) when NVMe reads are limited by a software token bucket and every retrieved block comes from NVMe (32\,GiB, 36 chunks, 99--165 chunks per point). Crosses: predictions of the analytical model registered before the runs; predicted zero stalls are drawn at the axis floor.}
\label{fig:ratelimit}
\end{figure}

Fig.~\ref{fig:ratelimit} lowers the read bandwidth.
On a\_yaw, prefetching keeps the stall at zero down to 0.25\,GB/s and starts to stall at 0.1\,GB/s (P50 75.9\,ms), whereas on-demand swap-in stalls 116, 454, 1130, and 2256\,ms at 1.0, 0.25, 0.1, and 0.05\,GB/s; b1\_ws gives the same values for on-demand swap-in to within 1\,ms.
The transition matches the model of Sec.~\ref{sec:char}: in steady state a chunk reads about one new int4 chunk, and near 0.1\,GB/s that read no longer fits in the compute window.
One case must be read with care.
On b1\_ws at 0.1 and 0.05\,GB/s, the P99 stall of prefetching (2621 and 6490\,ms) is higher than that of on-demand swap-in (2424 and 4844\,ms), although its median is lower.
Below the transition bandwidth, a prefetch that has fallen behind holds the read path when a burst of new blocks is needed, and the worst chunk then waits longer than with on-demand copies.
At 0.25\,GB/s this shows only in the P99 of b1\_ws (399\,ms against a median of 0.03\,ms), and it does not occur on a\_yaw.
\sys{} therefore targets the regime above the transition bandwidth, which an NVMe drive exceeds by a wide margin.

\textbf{SATA-class bandwidth.}
We emulate a SATA-class drive by limiting both reads and writes to 0.5\,GB/s with a software token bucket that allows bursts of up to 16\,MiB, and run a 2000-chunk round trip twice.
All ten registered predictions hold: no chunk stalls (the longest wait is 0.08\,ms), no write falls back to the synchronous path, the first NVMe write and read occur at chunks 228 and 1114 as in the unlimited run, and the median chunk time (1227.8 and 1226.3\,ms) stays within 0.1\% of the unlimited run (1226.7\,ms).
This is an emulation with a software limiter, not a measurement on a SATA drive; the SATA projection in Sec.~\ref{sec:char-model} uses vendor specifications.

\subsection{Cost of spilling to NVMe (Q3)}\label{sec:eval-write}

Table~\ref{tab:bgw} measures a 256-chunk run under 32\,GiB in three intervals.
Once the pool is full (the 225--240 interval in Table~\ref{tab:bgw}), writing each evicted block synchronously adds 102.0\,ms to every chunk, and 135.8\,ms on the return leg, where reads and writes overlap.
With background write-back the increments are 1.8 and 4.9\,ms, with zero stalls, no synchronous fallback, and a bit-identical video; write-back runs at about 1.5\,GB/s while busy.

\begin{table}[t]
\centering
\caption{Background write-back removes nearly all of the eviction cost. Chunk-time increment over the pool-not-full interval (32\,GiB, a\_yaw, 256 chunks, 3 rollouts).}
\label{tab:bgw}
\footnotesize
\resizebox{\columnwidth}{!}{%
\begin{tabular}{@{}lcc@{}}
\toprule
Interval & Synchronous eviction & Background write-back \\
\midrule
Chunks 3--225 (pool not full) & 0.0\,ms & 0.0\,ms \\
Chunks 225--240 (eviction) & +102.0\,ms & +1.8\,ms \\
Chunks 240--256 (revisits) & +135.8\,ms & +4.9\,ms \\
\bottomrule
\end{tabular}}
\end{table}

\subsection{A 37.5-minute session and long-gap recall (Q4)}\label{sec:eval-long}

\begin{figure}[t]
\centering
\includegraphics[width=\columnwidth]{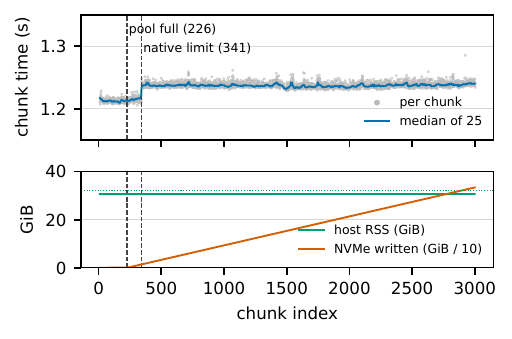}
\caption{Chunk time and host memory stay flat for 3000 chunks while KV spills to NVMe. One 3000-chunk session (37.5\,min of video) under 32\,GiB with all components enabled (1 rollout). Top: chunk time; the step at chunk 341 is the position re-mapping. Bottom: host RSS stays flat while KV spills to NVMe after the pool fills at chunk 226.}
\label{fig:session}
\end{figure}

\textbf{Continuous generation past both walls.}
Fig.~\ref{fig:session} shows one 3000-chunk session under 32\,GiB with all components enabled: int4 tiers, prefetching, background write-back, vectorized selection, the heap-based write path, and the hybrid position rule.
It produces 37.5 minutes of video in 63.2 minutes of wall-clock time (about 9.5\,fps over the session and 9.7\,fps at the median chunk time, below the 16\,fps playback rate; \sys{} does not make MG2 real-time).
The median chunk time is 1236.4\,ms (P99 1251.2\,ms), the longest stall is 0.12\,ms, and no chunk stalls for more than 0.5\,ms.
Host memory peaks at 30.7\,GiB, inside the 32\,GiB budget, and does not grow, while 334\,GiB are written to NVMe.
The chunk time steps up by 21.3\,ms at chunk 341, where the position re-mapping turns on (Sec.~\ref{sec:design-rope}), and is flat afterwards: the median of chunks 341--840 is 1237.4\,ms and that of the last 500 chunks is 1238.6\,ms.
This trajectory never returns to a spilled chunk, so the session exercises only the write path; reads from NVMe are measured next.

\textbf{Reading long-gap memory from NVMe.}
We run 2000-chunk round trips that walk out for 1000 chunks and return along the same path, so the return leg revisits chunks generated 21 to 1981 chunks earlier, most of them read from NVMe.
On MG2 in three scenes, no chunk that read from NVMe stalled for more than 0.5\,ms (1297 such chunks in the first scene and 648 in each of the others; the longest stall was 0.21\,ms), and every retrieved block was prefetched in time.
On LingBot-World~\cite{lingbot2} with round trips of 200 and 240 chunks (three repetitions each), every retrieval was a prefetch hit and the longest wait was 0.131\,ms.

\textbf{Does int4 tiered storage hurt long-gap recall?}
We compare \sys{} with the same system storing bf16 instead of int4 on 600-chunk round trips, with revisit gaps grouped as short ($<$100 chunks), mid (100--300), and long ($>$300).
We use three scenes, five seeds, and 13 revisit events per scene and gap tier; beyond chunk 341 both arms use the same position rule.
Table~\ref{tab:e6d} reports paired differences (int4 minus bf16) of two quantities: the LPIPS between a revisit frame and its first-visit frame, which measures consistency, and the MUSIQ score of the revisit frame, which measures image quality.
Consistency is not harmed: the LPIPS difference is $-$0.022 [$-$0.037, $-$0.008] in the mid tier and $-$0.026 [$-$0.043, $-$0.009] in the long tier; when seed variance is included, the mid-tier interval reaches zero.
Image quality on long gaps is about one point lower: the MUSIQ difference is $-$1.28 [$-$2.25, $-$0.36] in the long tier and $-$0.67 [$-$1.59, +0.17] in the mid tier.
Both differences stay within the tolerance registered before the runs ($|\Delta| \le 2$).
A lower LPIPS together with a lower MUSIQ is the direction of the copy effect discussed in Sec.~\ref{sec:analysis}, so we do not read the LPIPS difference as an improvement.

Five of the 15 scene--seed runs met the registered drift criterion (the two arms already differed by more than two MUSIQ points before the first revisit); excluding them gives a long-tier MUSIQ difference of $-$0.97 [$-$1.76, $-$0.17].
In scene 0003, which has no drifting run, all five seeds show a lower long-tier MUSIQ ($-$1.07 [$-$1.70, $-$0.45]).
Scene 0000 keeps only two seeds after the exclusion, so its per-scene numbers are weak.
Per-scene and per-seed values are listed in Appendix~\ref{app:e6d}.
On the short-horizon sealed test set below, MUSIQ is unchanged; the loss appears only for long gaps beyond the native horizon.

\begin{table}[t]
\centering
\caption{Int4 storage does not harm revisit consistency and lowers long-gap MUSIQ by about one point. Int4 minus bf16 tiered storage on long-gap revisits (3 scenes, 5 seeds, 39 events per tier; 95\% event-level bootstrap intervals). Each event is first averaged over the seeds kept, so the three scenes weigh equally; after the drift exclusion scenes keep 2, 5, and 3 seeds, and the values differ from a plain mean over the 10 runs in Appendix~\ref{app:e6d}. $\Delta$LPIPS $<0$: revisit closer to first visit. $\Delta$MUSIQ $<0$: lower image quality.}
\label{tab:e6d}
\footnotesize
\setlength{\tabcolsep}{2.5pt}
\resizebox{\columnwidth}{!}{%
\begin{tabular}{@{}llcc@{}}
\toprule
Runs & Gap & $\Delta$LPIPS & $\Delta$MUSIQ \\
\midrule
All 15 (main) & short & $-$0.005 [$-$0.015, +0.005] & $-$0.04 [$-$0.36, +0.28] \\
 & mid & $-$0.022 [$-$0.037, $-$0.008] & $-$0.67 [$-$1.59, +0.17] \\
 & long & $-$0.026 [$-$0.043, $-$0.009] & $-$1.28 [$-$2.25, $-$0.36] \\
\midrule
No drift (10) & short & $-$0.015 [$-$0.024, $-$0.007] & +0.15 [$-$0.41, +0.70] \\
 & mid & $-$0.036 [$-$0.058, $-$0.015] & $-$0.86 [$-$2.37, +0.60] \\
 & long & $-$0.021 [$-$0.035, $-$0.006] & $-$0.97 [$-$1.76, $-$0.17] \\
\bottomrule
\end{tabular}}
\end{table}

\subsection{Quality and a second model (Q5)}\label{sec:eval-quality}

\textbf{Sealed test set.}
To keep method development from leaking into the quality claim, we froze a test set of four new scenes and two trajectories (eight units, seeds 5--9) and evaluated the final configuration on it once.
We measure the fraction of the revisit benefit of full-precision memory that is retained, $(L_{\mathrm{sw}}-L_x)/(L_{\mathrm{sw}}-L_{\mathrm{bf16}})$, where $L$ is the revisit LPIPS and sw is the sliding window.
With int4 storage, retention is 105\% [104\%, 106\%] and MUSIQ changes by $-$0.06 [$-$0.19, +0.07].
Retention slightly above 100\% is the direction of the copy effect (Sec.~\ref{sec:analysis}), not better memory, and the MUSIQ interval includes zero; we therefore find no measurable quality loss from int4 on this set.

\textbf{Second model.}
LingBot-World~\cite{lingbot2} (1.3B variant; licensed CC BY-NC-SA) was not trained with retrieval memory, and in our tests retrieval did not improve its revisit consistency, so we report system metrics only (Table~\ref{tab:lingbot}).
Its chunk length differs from that of MG2; one chunk of its KV is 281.7\,MiB in int4.
Under 32 and 64\,GiB the first NVMe write occurs at chunks 102 and 218, and the same pool holds about 3.8$\times$ more int4 than bf16 chunks.
With prefetching, all 750 retrievals are hits and no chunk waits more than 0.131\,ms across both repetitions (the table shows the first).
With on-demand swap-in, every chunk with retrieval waits: 37.7\,ms at the median and up to 375\,ms.
Retrieval itself makes a chunk 35.5\% slower than the sliding window, because attention covers 12 more frames.

\begin{table}[t]
\centering
\caption{LingBot-World, 256 chunks (its native limit), system metrics only (first of two repetitions).}
\label{tab:lingbot}
\footnotesize
\setlength{\tabcolsep}{2.5pt}
\resizebox{\columnwidth}{!}{%
\begin{tabular}{@{}lcccc@{}}
\toprule
Design (budget) & Chunk (ms) & Wait P50 / max (ms) & 1st write & 1st read \\
\midrule
Sliding window & 2576 & -- & -- & -- \\
On-demand swap-in (32\,GiB) & 3488 & 37.7 / 375.4 & 102 & 179 \\
\sys{} (32\,GiB) & 3491 & 0.017 / 0.103 & 102 & 178 \\
\sys{} (64\,GiB) & 3496 & 0.017 / 0.107 & 218 & 236 \\
\bottomrule
\end{tabular}}
\end{table}

\section{Analysis and Findings}\label{sec:analysis}

This section reports findings that shaped the design, including approaches that did not work.
Quality numbers here come from a development set of six scenes and two trajectories (12 units, seeds 5--9) on which we tuned methods, so they are weaker evidence than the sealed set of Sec.~\ref{sec:eval-quality}.
Retention is defined as in Sec.~\ref{sec:eval-quality}.

\textbf{At equal bytes, quantization beats head sparsity.}
Before settling on int4, we stored KV only for a subset of attention heads, a natural idea because heads differ in how much they attend to distant context.
At about a quarter of the bytes, keeping all heads in int4 (26.6\% of bf16) retains 105\% [102\%, 108\%] of the memory benefit, while keeping 25\% of the heads in bf16 retains 56\% [51\%, 61\%], a difference of 48.1 points [41.0, 55.3] that holds in all 12 units.
Choosing the heads by an online ranking does not help (58\%), even though head rankings are stable across scenes and events; a random half of the heads in int4 (13.3\% of the bytes) retains 70\%.
Fidelity is better spent on every head at lower precision than on a few heads at full precision.

\textbf{Memory value concentrates on revisits.}
If heads are restored only when a revisit is predicted, retention with a random 25\% of heads rises from 56\% to 81\%.
Most of the benefit of memory is realized on the few chunks that return to a known place.
We tried to exploit this by reading remote memory only when a revisit is predicted.
It preserved 82\% [78\%, 86\%] of the benefit on pose-matched revisits but raised throughput by only 1.6--3.9\% on our trajectories, because the staging reuse of Sec.~\ref{sec:design-prefetch} already moves only about one chunk per step; we did not adopt it.
Its trigger also fires more often than needed on long pure rotations, because it normalizes the pose distance by the largest distance in the candidate pool.

\textbf{Bit width and the copy effect.}
Below 4 bits, revisit LPIPS keeps falling while image quality drops: int3 retains 128\% with a MUSIQ change of $-$0.90 [$-$1.50, $-$0.32], and symmetric int2 retains 120\% with $-$4.05 [$-$6.44, $-$2.01].
Retention above 100\% does not mean better memory.
Coarser keys and values make the revisited frame copy its first visit more closely, which lowers LPIPS, while texture quality degrades.
Revisit LPIPS must therefore be read together with a quality score; this is why Sec.~\ref{sec:eval-long} reports LPIPS and MUSIQ side by side.
Correcting the attention bias of quantized keys~\cite{qksa} restores 2-bit quality on our development set (asymmetric int2 with the correction: 17.2\% of the bytes, 102\% retention, MUSIQ $+$0.30), but in our implementation it lengthens attention by 63--124\% in microbenchmarks on synthetic tensors.
\sys{} is agnostic to the bit width; int4 is the default, and corrected int2 is an optional higher-compression setting.

\textbf{The model's own memory weakens with distance, and the benefit of position re-mapping depends on the scene.}
In the 2000-chunk round trips of Sec.~\ref{sec:eval-long}, we measure the fraction of the sliding window's revisit inconsistency that memory removes, $\rho = 1 - L_x/L_{\mathrm{sw}}$, for short (21--93 chunks), mid, and long (1021--1981 chunks) revisit gaps.
In scene 0000 (two seeds), the long-to-short ratio of $\rho$ is 0.96 [0.76, 1.15] with the extended position table alone and 0.86 [0.76, 0.97] with re-mapping; in scene 0003 (one seed) it is 0.73 and 0.78.
In scene 0011 the sliding window itself barely changes on revisits (LPIPS 0.05 on short gaps), so $\rho$ is undefined in practice and we report absolute LPIPS instead: 0.338, 0.311, and 0.360 without re-mapping and 0.271, 0.270, and 0.314 with it for short, mid, and long gaps.
Gap and position in the session are confounded on these trajectories, because long gaps occur late in the return leg.

In paired comparisons with one seed per scene (unlike the two-seed ratio above), re-mapping lowers revisit LPIPS by 0.15--0.28 in scene 0000 at all gaps and by 0.04--0.07 in scene 0011 at short and mid gaps (its long-gap interval includes zero), and has no measurable effect in scene 0003.
Placing retrieved memory too close (6 frames before the current chunk) or too far (96 frames) is worse than our setting of 30 frames.
We therefore treat re-mapping as an optional component whose benefit depends on the scene.

\textbf{Long-range texture drift is not caused by quantization.}
In long sessions, generation with retrieval memory develops a repeated diagonal texture after about 100 chunks.
A control that stores bf16 instead of int4 on the same trajectory and noise shows the same texture: over chunks 100--599 the texture energy ratio is 0.97 and the MUSIQ difference is $-$0.11.
The sliding window does not show this texture, but its frames leave the style of the starting image, so it is not a clean reference.
We attribute the drift to memory-conditioned generation itself, not to the precision of the stored KV.

\section{Related Work}\label{sec:related}

\textbf{Memory for interactive world models.}
Retrieval memory keeps world-model generations consistent on revisits.
WorldKV~\cite{worldkv} retrieves evicted KV chunks by camera pose or action without training and keeps them in GPU or CPU memory; its authors note that host-to-device transfer latency limits the CPU-offloaded variant.
ReWorld~\cite{reworld} trains a model with a bounded bank of landmark chunks held in pinned host memory and prefetches the landmarks nearest to the current pose on a separate stream, overlapped with denoising.
OmniMem~\cite{omnimem} retrieves history per attention head for long video generation, keeps a small LRU cache on the GPU, and waits for blocks that must be reloaded from pinned host memory.
WorldTrace~\cite{worldtrace} keeps evicted keys in host memory and makes memory addressable beyond the native context without training.
WorldPlay~\cite{worldplay} retrieves past frames by field-of-view overlap and camera distance, reassigns their temporal positions, and recomputes the memory KV for every chunk.
Velox~\cite{velox} reduces attention compute over resident KV with geometric routing and keeps KV on the GPU.
Inferix~\cite{inferix}, an inference engine for world simulation, supports offloading KV to main memory.
These works decide \emph{what} a chunk should attend to, and some move memory to host DRAM; \sys{} decides \emph{where} that memory lives and \emph{when} it moves, and can run below any pose-keyed retrieval policy.
To our knowledge, none of them places world-model KV on an NVMe tier; Table~\ref{tab:related} summarizes these systems.
Velox-style routing and WorldKV-style token compression reduce compute or bytes per chunk and could be combined with \sys{}; we have not tested such combinations.

\textbf{KV quantization for video and world models.}
KIVI~\cite{kivi} quantizes keys per channel and values per token.
QuantWM~\cite{quantwm}, Quant VideoGen~\cite{qvg}, and TempoKV~\cite{tempokv} quantize the KV of video and world models to 2--5 bits, and Tuncer et al.~\cite{qksa} correct the attention bias of quantized keys.
These methods reduce bytes per chunk by a constant factor and thus delay the capacity wall; \sys{} uses a KIVI-style int4 quantizer as a component and adds the tiers and prefetching that handle what still does not fit.

\textbf{KV offloading and tiering for LLMs.}
vLLM~\cite{vllm} manages KV memory in pages and can swap it to CPU memory.
FlexGen~\cite{flexgen} offloads weights and KV across GPU, CPU, and disk for throughput-oriented batch inference on one GPU, with I/O overlapped with compute in a fixed schedule.
InfiniGen~\cite{infinigen} speculates which KV entries the next layer needs and prefetches them from CPU memory.
CachedAttention~\cite{cachedattention} caches conversation KV in host memory and disk, preloads it layer by layer, saves it asynchronously, and decouples positional encoding from cached KV.
PowerInfer~\cite{powerinfer} serves LLMs on a consumer-grade GPU.
Quest~\cite{quest} shows that which KV entries matter depends on the query.
The LLM setting differs from ours in two ways that change the design: a decoding step is short, so moving KV is bandwidth-bound, and a selective access set is known only after the query is computed.
In a world model a chunk computes for about a second and its access set follows from the camera pose, which makes zero-stall prefetching from NVMe possible on a consumer machine.

\textbf{Long video beyond the native horizon.}
StreamingLLM~\cite{streamingllm} keeps attention-sink tokens and assigns positions within the cache.
Rolling Forcing~\cite{rollingforcing} re-applies RoPE to cached keys at effective indices so that they keep a fixed relative position to the frames being denoised, and Deep Forcing~\cite{deepforcing} shifts the temporal RoPE phase of sink tokens to the current timeline.
WorldPlay applies a related reframing to retrieved memory in an interactive world model.
\sys{}'s position handling follows this practice; we apply it to pose-retrieved long-term memory that may be read back from NVMe and enable it only beyond the native horizon, so that the first 341 chunks stay bit-identical.

\begin{table*}[t]
\centering
\caption{Positioning. ``--'': not applicable or not stated in the paper. Entries summarize each paper's own description (see Sec.~\ref{sec:related}).}
\label{tab:related}
\footnotesize\setlength{\tabcolsep}{4pt}
\begin{tabular}{@{}lllllll@{}}
\toprule
Work & Workload & Where KV lives & Prefetch signal & On a miss & Training-free & Beyond native pos. \\
\midrule
WorldKV~\cite{worldkv} & world model & GPU / CPU & -- & -- & yes & -- \\
ReWorld~\cite{reworld} & world model & GPU + host DRAM (bank $\le$30) & current pose & -- & no & -- \\
OmniMem~\cite{omnimem} & long video & GPU + host DRAM & -- (reactive) & waits & no & -- \\
WorldPlay~\cite{worldplay} & world model & frames; KV recomputed per chunk & -- & -- & no & yes (reframing) \\
WorldTrace~\cite{worldtrace} & world model & GPU + host DRAM (keys) & -- & -- & yes & yes \\
Velox~\cite{velox} & world model & GPU & -- & -- & yes & -- \\
Inferix~\cite{inferix} & world simulation & GPU + main memory & -- & -- & -- & -- \\
FlexGen~\cite{flexgen} & LLM (batch) & GPU / CPU / disk & fixed schedule & -- & yes & -- \\
InfiniGen~\cite{infinigen} & LLM & GPU + CPU & speculation & -- & yes & -- \\
CachedAttention~\cite{cachedattention} & LLM & GPU / DRAM / disk & scheduler queue & -- & yes & decoupled PE \\
\textbf{\sys{}} & world model & GPU / DRAM / \textbf{NVMe} (int4) & extrapolated pose & stale set, no wait & yes & yes (hybrid) \\
\bottomrule
\end{tabular}
\end{table*}

\section{Limitations and Conclusion}\label{sec:concl}

\textbf{Limitations.}
Our evidence has the following scope.
\begin{itemize}
\item \emph{One machine, one drive.} All timing comes from one RTX~4090 with one NVMe SSD; slower drives are emulated with a software token bucket, including the SATA-class case.
\item \emph{One scene and one seed for timing.} System measurements use one starting image and one seed per configuration (with up to five rollouts), so we do not report confidence intervals for them.
\item \emph{Quality on one model.} Image-quality and consistency results are for MG2 only; on LingBot-World retrieval did not improve consistency, so we report system metrics only for it.
\item \emph{Long-gap image quality.} int4 tiered storage lowers MUSIQ on long-gap revisits by about one point ($-$1.28 [$-$2.25, $-$0.36]), within the registered tolerance; scene 0000 keeps only two seeds after drifting runs are excluded.
\item \emph{Long-range texture drift.} Memory-conditioned generation develops a repeated texture after about 100 chunks, independently of KV precision (Sec.~\ref{sec:analysis}).
\item \emph{Model accuracy.} Three held-out predictions miss (near the failure bandwidth, LingBot at 0.075\,GB/s, and LingBot read-back volume), and int8 was emulated by byte accounting only.
\item \emph{Speed.} At about 9.5 frames per second, \sys{} generates video continuously but not in real time.
\end{itemize}

\textbf{Conclusion.}
On a single consumer machine, long sessions of a memory-augmented world model stop at two walls: host memory runs out after 45 seconds of video, and the model's positions run out after 4.3 minutes.
The memory is capacity-bound rather than bandwidth-bound, and its access set is known before compute, so an analytical model can predict where each budget spills and what bandwidth prefetching needs.
\sys{} follows that model: it stores int4 KV in a GPU, DRAM, and NVMe hierarchy, prefetches from the extrapolated pose without ever waiting, writes back in the background, and handles positions at inference time.
One RTX~4090 with 32\,GiB of host memory and one NVMe drive then generates 37.5 minutes of video continuously, with zero stalls and constant host memory.

\section*{Use of AI}
Large language models (Anthropic's Claude) were used to draft and revise the text of this paper, to write the figure-generation scripts, and to write and run parts of the experiment code under the authors' direction. All experimental designs were registered before the runs, every number in the paper was traced to a result file, and the authors reviewed the full text and take responsibility for its content.

\bibliographystyle{IEEEtran}
\bibliography{refs}

\appendices
\section{Per-scene and per-seed results for int4 versus bf16 long-gap recall}\label{app:e6d}
Table~\ref{tab:e6d-seed} lists, for each scene and seed, the paired differences (int4 minus bf16) averaged over the 13 revisit events of each gap tier, and the drift flag. A run is flagged when the two arms differ by more than two MUSIQ points on the outbound leg (chunks 3--299) or on the first-visit frames, that is, before any revisit.
\begin{table}[t]
\centering
\caption{Int4 minus bf16 per scene and seed (event means, no intervals). D: drift flag.}
\label{tab:e6d-seed}
\footnotesize
\resizebox{\columnwidth}{!}{%
\begin{tabular}{@{}llc rrr rrr@{}}
\toprule
Scene & Seed & D & \multicolumn{3}{c}{$\Delta$LPIPS (short / mid / long)} & \multicolumn{3}{c}{$\Delta$MUSIQ (short / mid / long)} \\
\midrule
0000 & 5 &  & +0.016 & $-$0.116 & $-$0.089 & $-$0.54 & $-$2.56 & $-$1.73 \\
0000 & 6 &  & $-$0.062 & $-$0.021 & +0.041 & $-$0.77 & $-$3.22 & $-$0.99 \\
0000 & 7 & yes & +0.128 & +0.076 & +0.035 & $-$1.37 & $-$7.49 & +2.33 \\
0000 & 8 & yes & +0.035 & $-$0.029 & +0.014 & $-$0.80 & +1.52 & $-$11.03 \\
0000 & 9 & yes & $-$0.068 & $-$0.033 & $-$0.065 & +0.35 & +3.35 & +0.35 \\
\midrule
0003 & 5 &  & +0.080 & +0.005 & $-$0.027 & +2.98 & +1.74 & $-$1.93 \\
0003 & 6 &  & +0.010 & $-$0.031 & $-$0.047 & +0.58 & $-$3.13 & $-$1.47 \\
0003 & 7 &  & $-$0.067 & $-$0.071 & +0.008 & $-$1.16 & $-$0.34 & $-$0.46 \\
0003 & 8 &  & $-$0.005 & +0.001 & $-$0.021 & $-$1.20 & $-$0.98 & $-$1.30 \\
0003 & 9 &  & $-$0.037 & $-$0.010 & $-$0.022 & $-$1.64 & +0.08 & $-$0.19 \\
\midrule
0011 & 5 &  & $-$0.031 & +0.010 & +0.002 & +1.07 & +2.71 & +0.31 \\
0011 & 6 &  & $-$0.003 & $-$0.018 & $-$0.048 & +1.28 & $-$1.64 & $-$2.00 \\
0011 & 7 & yes & $-$0.044 & $-$0.069 & $-$0.030 & $-$2.84 & $-$2.70 & $-$0.02 \\
0011 & 8 & yes & $-$0.006 & +0.021 & $-$0.135 & +2.31 & +1.24 & $-$1.41 \\
0011 & 9 &  & $-$0.021 & $-$0.044 & $-$0.007 & +1.19 & +1.39 & +0.29 \\
\bottomrule
\end{tabular}}
\end{table}

\end{document}